\documentclass[sn-mathphys]{sn-jnl}% Math and Physical Sciences Reference Style
\begin{document}

\title[Amphiphilic diblock copolymers as functional surfaces]{Amphiphilic diblock copolymers as functional surfaces for protein chromatography}


\author*[1]{\fnm{Raghu} \sur{K. Moorthy}}\email{raghukmoorthy@iitb.ac.in}

\author[1]{\fnm{Serena} \sur{D'Souza}}

\author[1]{\fnm{P.} \sur{Sunthar}}

\author[1]{\fnm{Santosh} \sur{B. Noronha}}

\affil*[1]{\orgdiv{Department of Chemical Engineering}, \orgname{Indian Institute of Technology Bombay}, \orgaddress{\street{Powai}, \city{Mumbai}, \postcode{400076}, \state{Maharashtra}, \country{India}}}


\abstract{Stationary phase plays a crucial role in the operation of a protein chromatography column. Conventional resins composed of acrylic polymers and their derivatives contribute to heterogeneity of the packing of stationary phase inside these columns. Alternative polymer combinations through customized surface functionalization schemes which consist of multiple steps using static coating techniques are well known. In comparison, it is hypothesized that a single-step scheme is sufficient to obtain porous adsorbents as stationary phase for tuning surface morphology and protein immobilization. To overcome the  challenge of heterogeneous packing and ease of fabrication at a laboratory scale, a change in the form factor of separation materials has been proposed in the form of functional copolymer surfaces. In the present work, an amphiphilic, block copolymer, poly(methyl methacrylate-co-methacrylic acid) has been chosen and fully characterized for its potential usage in protein chromatography. Hydrophilicity of the acrylic copolymer and abundance of carboxyl groups inherently on the copolymer surface have been successfully demonstrated through contact angle measurements, Fourier transform infrared (FTIR) and X-ray photoelectron spectroscopy (XPS) studies. Morphological studies indicate presence of a microporous region (nearly 1 to 1.5 $\mu$m pore size) that could be beneficial as a cation exchange media as part of the stationary phase in protein chromatography.
}


\maketitle

\section{Introduction}\label{sec1}

Functional surfaces that cater to the needs of biocompatibility are in demand from a wide spectrum of users, ranging from bioanalytics to biopharmaceutical industries (\cite{Barrett2005}, \cite{Bratek-Skicki2019}). To promote immobilization of biomolecules, tailor-made surfaces are leveraged using different polymeric materials. For example, plastics are modified to specifically adsorb proteins \cite{Kang1993}. Functional groups commonly used in ionic interactions possess characteristic charges and determine the adsorption rates of various types of proteins including those which have been subjected to post-translational modifications. Citing rapid prototyping and relatively lower costs, polymers have evolved as reliable materials over the past two decades (\cite{Liu2000}, \cite{Becker2002}, \cite{Lee2010}, \cite{Ouimet2022}). 

Diblock copolymers are ordered polymer blocks with two or more different monomers covalently bonded to provide synergistic surface and chemical properties \cite{Kenney1968}. Wide variety of such copolymers consist of acrylic derivatives as one of the monomers due to their thermoplastic behaviour and being chemically inert. Acrylic copolymer is one among them with methacrylic acid providing the chemical functionality (\cite{Hogt1985}, \cite{Saunders1997}). From existing literature, a comparative tabulation has been shown to emphasize the physical properties associated with polydimethylsiloxane (PDMS) and acrylic copolymer (ACP) over the common coating substrates or supports (Table \ref{tab:substratecomparison}) (\cite{Pulker1999}, \cite{Sigma2018}). Physical interactions mainly through hydrogen bonding contribute towards the attractive forces between the electronegative oxygen atom (from carboxylate groups of methyl methacrylate part of the acrylic copolymer) and the hydrogen atom (from hydroxyl groups of alkali-treated polydimethylsiloxane (PDMS)) \cite{Liu2000}. Hydrogen atom in the carboxylate group of methacrylic acid part of the acrylic copolymer chain approaches hydroxide ions (Equation \ref{eqn:rxnscheme} where R and R' are different alkyl chains from ACP and PDMS respectively). Thus, it leaves the highly electronegative oxygen atom bonded to -$\mathrm{sp}^2$ carbon in an ionized state to form ${\mathrm{COO}}^{-}$ groups \cite{Hogt1985}. These negatively charged functional groups are beneficial as cation exchange media in protein separations (\cite{Carta2020}, \cite{Hasan2018}, \cite{Phan2015}). Various alternate combinations of functional polymer surfaces in accordance to the above requirements have been developed (\cite{Hjerten1985}, \cite{Hoek2010}, \cite{Hosseini2014}, \cite{Hosseini2014a}).     
\begin{equation}
	\mathrm{R-COOH } + \mathrm{ } ^{-}\mathrm{OH-R'} \mathrm{ } \longrightarrow \mathrm{ R-COO}^{-} \mathrm{ } + \mathrm{ H---HO-R'}
	\label{eqn:rxnscheme}
\end{equation}
In the present work, formation of large pores using acrylic copolymer is investigated for its suitability as a porous stationary phase in protein chromatography. It is hypothesized that implementing single-step functionalization scheme leads to functional surfaces that promote protein adsorption. Further, a change in the form factor of separation materials (Table \ref{tab:functionalizationliterature}) from functionalized surfaces \cite{Ouimet2022} to functional copolymer surfaces inside the separation column is likely to ease the fabrication challenges at a laboratory scale and provide improved control over surface morphology in presence of necessary functional groups.

\section{Experimental}\label{sec2}

\subsection{Chemicals and reagents}\label{subsec1}
Polydimethyldisiloxane (PDMS) elastomer kit (Sylgard 184) was supplied from Dow Corning Incorporated, USA. Monosodium dihydrogen phosphate (Cat No. 59443) and disodium hydrogen phosphate (Cat. No. 21669) (all anhydrous) (Sisco Research Laboratories, India) for phosphate buffer preparation, Bradford reagent (Cat. No. ML-106, HiMedia Laboratories, India) for protein detection were purchased. Polymethyl methacrylate (PMMA) (Cat. No. 182230) and poly(methyl methacrylate-co-methacrylic acid) (85:15 monomer ratio) (Cat. No. 376914) (hereafter, referred to as acrylic copolymer (ACP)) (both from Sigma, USA) were procured. These chemicals were used as received without further modification. Proteins used in this work include bovine serum albumin (BSA) (Cat. No. 85171) (Sisco Research Laboratories, India) and human serum albumin (HSA) (Cat. No. A1653) (both from Sigma, USA). Tetrahydrofuran (THF) was the organic solvent used for ACP dissolution. 5 M sodium hydroxide (NaOH) and 1 M hydrochloric acid (HCl) solutions were prepared and used in polymer coating procedure \cite{Hoek2010} and for adjustment of pH during preparation of phosphate buffers respectively. The prepared solutions were filtered and sonicated before use. All analytical grade chemicals or reagents unless stated otherwise were purchased from Merck Laboratories, India. Deionized (DI) water (Millipore, USA) has been utilized throughout the experiments.

\subsection{Preparation of acrylic copolymer surface}\label{subsec2}
\begin{figure}[tbp]
	\includegraphics[width=0.7\textwidth]{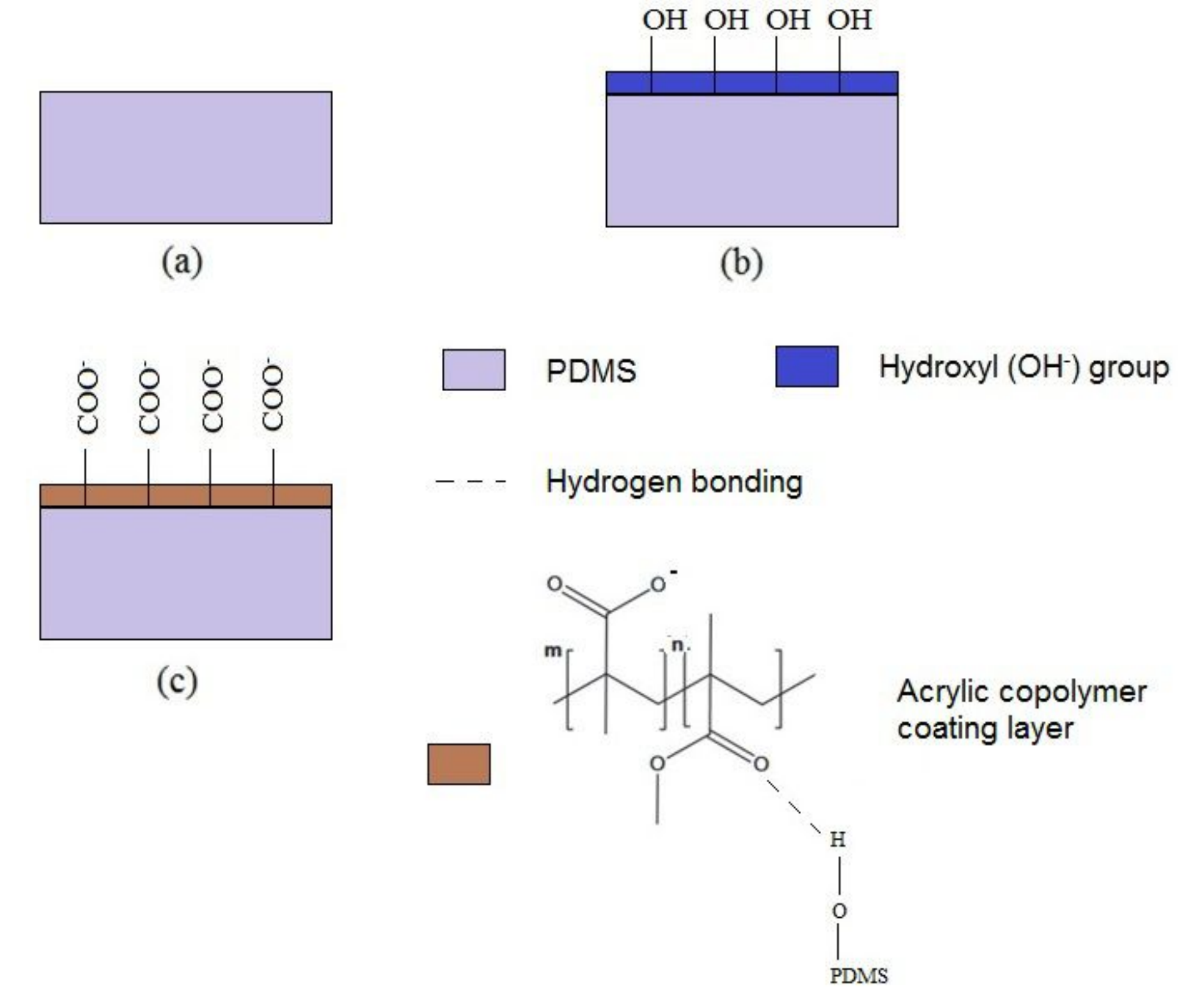}
	\centering
	\caption{Schematic diagram of polymer surface functionalization - Base substrate: polydimethylsiloxane (PDMS), coated wall material: poly(methyl methacrylate-co-methacrylic acid) [poly(MMA-co-MAA)] (acrylic copolymer or ACP)
		(a) uncoated PDMS  (b) alkali-treated (blue-shaded) hydrophobic PDMS  (c) ACP-coated PDMS surface (brown-shaded) exposed with negatively charged carboxyl groups}
	\label{fig:copolymersurfacechemistryschematic}
\end{figure}

PDMS surfaces with acrylic copolymer were prepared using a modified static coating method \cite{Pulker1999}.	The surface functionalization was achieved as shown in Figure \ref{fig:copolymersurfacechemistryschematic}. PDMS layer was formulated using silicone elastomer and curing agent mixed in the ratio of 10:1. The PDMS solution was poured slowly into a glass petri dish and subjected to curing process at temperature 75$^{o}$C for 2 hours. Flat PDMS blocks with a rectangular area of 20 x 10 mm each were utilized as coating support. Different concentrations (5, 10 and 15\% w/v) of acrylic copolymer in tetrahydrofuran (THF) were to be obtained through a three-step procedure that involved alkali treatment, cleaning and drying. PDMS blocks were rinsed with 5 M sodium hydroxide for 10 to 15 minutes \cite{Hoek2010} and washed with deionized water. These PDMS surfaces were placed on a borosilicate glass support and ~100 $\mu$l acrylic copolymer solution was drop-dispensed onto it. The acrylic copolymer solution was spin-coated at 500 rpm for 1 minute to achieve coated film on the top surface of PDMS block. To assist in solvent evaporation, the above coated samples were left for solvent drying at room temperature (minimum 1 hour) before further analysis.
\label{PDMSblockpreparation}

\subsection{Contact angle measurements}\label{subsec3}
Water contact angle for the coated PDMS samples were measured using sessile drop method (GBX Digidrop contact angle meter, France). A water drop of 2 $\mu$l was placed at a central point and at four corners over the coated area and the average contact angle was calculated along with their standard deviation (n=3).    

\subsection{Scanning electron microscopy (SEM)}\label{subsec4}
Surface imaging of the polymer samples were performed using field emission scanning electron microscope (FESEM, JEOL JSM-7600F, Japan) at accelerating voltage of 5 or 10 kV. The polymer surface of interest was coated with a thin film of platinum or iridium to negate surface charging (film thickness of ~10 nm). The images have been captured at a working distance ranging between 5 to 15 mm away from the sample surface.

\subsection{Fourier transform infrared (FTIR) spectroscopy}\label{subsec5}
To confirm the presence of carboxyl groups, the routine FTIR spectroscopy studies were carried out in the wave number range 1000-4000 cm$^{-1}$. The KBr pellet method was adopted to obtain sample pellets for further analysis. 50 to 60 mg potassium bromide (KBr) was mixed, and ground with polymer sample to obtain 1-2 mg of final sample. It was transferred to the mold and subjected to high pressure (~100 kgf/cm$^{2}$) to form a pellet of thickness less than 1 mm. Infrared light beam was used for further analysis of the pellet (3000 Hyperion microscope with Vortex 80 FTIR system, Bruker, Germany). Sample spectrum has been reported after necessary baseline and straight line corrections. Further, coated PDMS blocks (rectangular area of 20 x 10 mm each) were used to analyze the distribution of carboxyl groups on the coated layer. In order to cover a wider region of interest, FTIR spectra was collected with focal plane array (FPA) detector (scanned area:128 x 128 pixels) (3000 Hyperion microscope with Vortex 80 FTIR system, Bruker, Germany). The target peak was expected to be in the wavelength range of 1600 to 1850 cm$^{-1}$ (for carboxyl groups) as in previous literature \cite{Duan2008}. Sample spectrum was reported after necessary baseline and straight line corrections using barium fluoride as standard reference chemical that did not contain carboxyl groups.

\subsection{X-ray photoelectron spectroscopy (XPS)}\label{subsec6}
The XPS analysis mode has been carried out using monochromatic (AlK$\alpha$) 600 W X-ray source with an energy resolution of 0.7 eV and emission angle of 0$^{o}$ (AXIS Supra, Kratos Analytical, Shimadzu Group, UK). The analysis depth is ~10 nm for polymer samples of spot size less than 5 mm. Spectral data for elemental carbon ($^{12}$C) have been reported for different acrylic copolymer concentrations and corresponding chemical bonds were confirmed.

\subsection{Bradford assay at microliter capacity}\label{subsec7}
Bovine serum albumin was considered for protein calibration standards using Bradford assay (\cite{Bradford1976}, \cite{Grintzalis2015}). The optical detection was performed using a microtiter plate reader (Multiskan GO, ThermoFisher Scientific, USA) at 595 nm using 150 $\mu$l detection volume (30 $\mu$l of protein and 120 $\mu$l of Bradford reagent) to obtain a pronounced response signal. A 1:4 protein-dye volume ratio lowered the interference due to background signal of Bradford reagent.

\subsection{Batch protein adsorption studies}\label{subsec8}
Protein adsorption with varying feed concentration was performed using both uncoated and coated, PDMS blocks. Binding capacity of the acrylic copolymer (or adsorbent) was quantified through these batch studies. Coated PDMS blocks previously obtained (in Section \ref{PDMSblockpreparation}) through static coating technique was utilized in this procedure. Protein in 20 mM phosphate buffer, pH 6.4 was used as feed in the adsorption studies. The above sample (150 $\mu$l) was loaded onto the functional surface in each block at different protein concentrations, that varied between 100 and 1000 $\mu$g ml$^{-1}$. Uncoated PDMS blocks were taken as control sample for comparison purposes. All of the above blocks were subjected to slow shaking at 250 rpm for 90 minutes (Cole-Parmer, India). To calculate the binding capacity of ACP (as in Equation \ref{eqn:bindingcapacity}), 30 $\mu$l of the buffer solution was collected from top surface of the block in each case with internal duplicate samples and the unbound protein concentration was estimated with Bradford assay (n=3).
\begin{equation}
	\textnormal{Binding capacity} \left(\textnormal{in }\%\right) = \frac{\textnormal{Weight of adsorbed protein}*100}{\textnormal{Weight of feed protein}}
	\label{eqn:bindingcapacity}
\end{equation}

\section{Results and Discussion}\label{sec3}

\subsection{Surface morphology}\label{subsec9}
\begin{figure}[tbp]
	\includegraphics[width=0.7\textwidth]{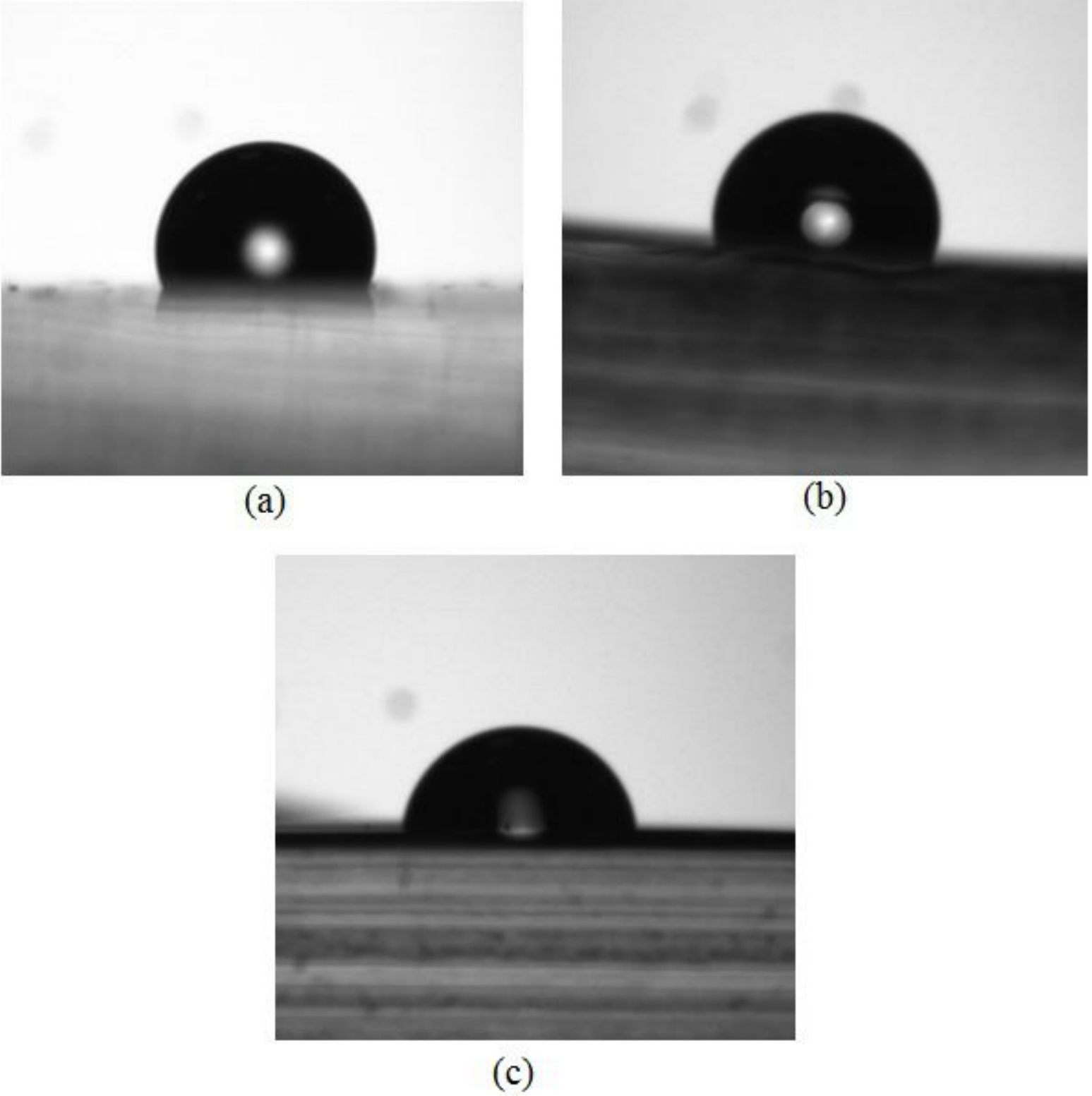}
	\centering
	\caption{Water contact angle of: (a) uncoated PDMS (b) 5\%(w/v) acrylic copolymer on PDMS (c) 10\%(w/v) acrylic copolymer on PDMS}
	\label{fig:copolymercontactangle1}
\end{figure}

\begin{figure}[tbp]
	\includegraphics[width=0.7\textwidth]{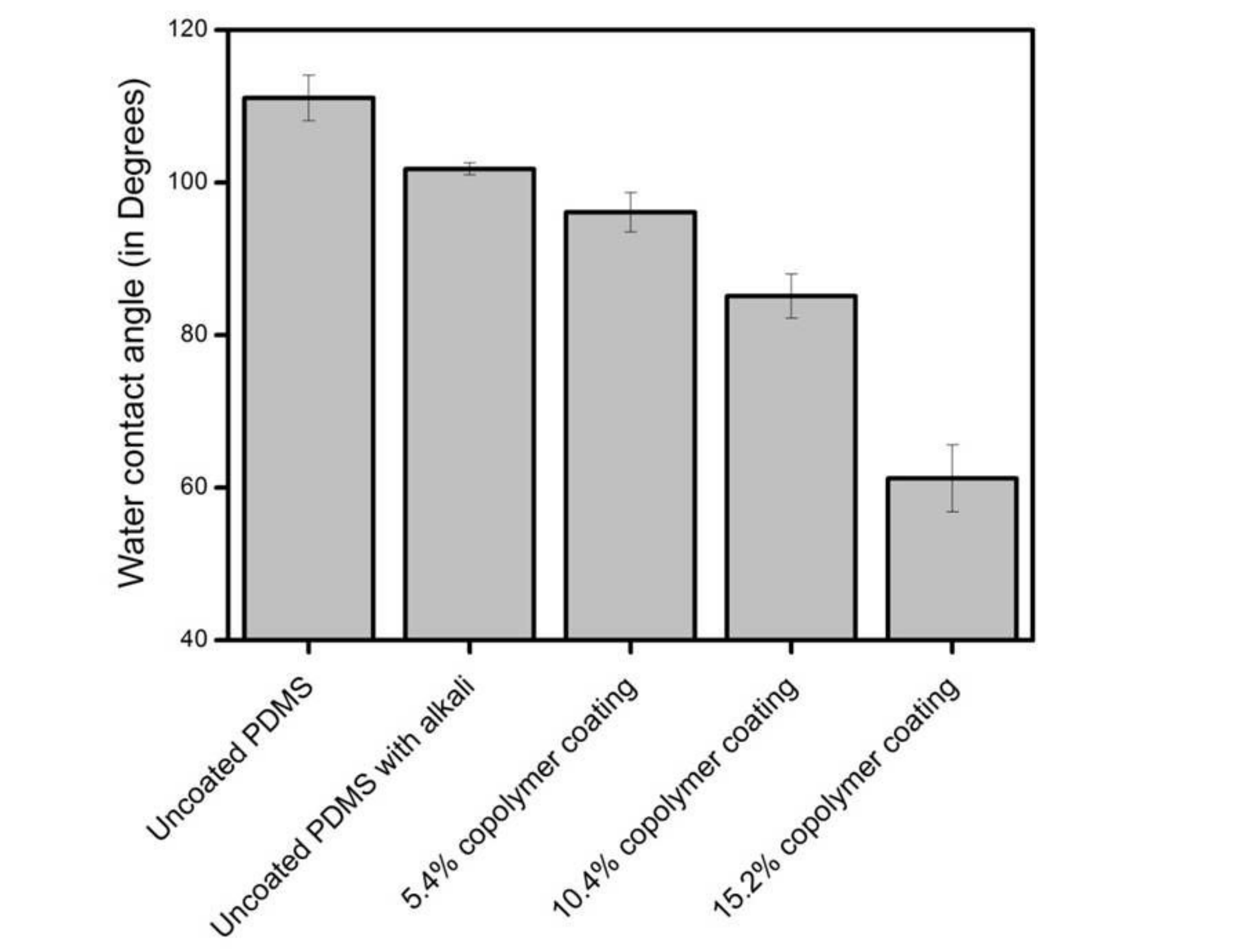}
	\centering
	\caption{Graphical representation of water contact angles of uncoated PDMS and acrylic copolymer-modified PDMS surfaces}
	\label{fig:copolymercontactangle2}
\end{figure}

\begin{figure}[tbp]
	\includegraphics[width=0.7\textwidth]{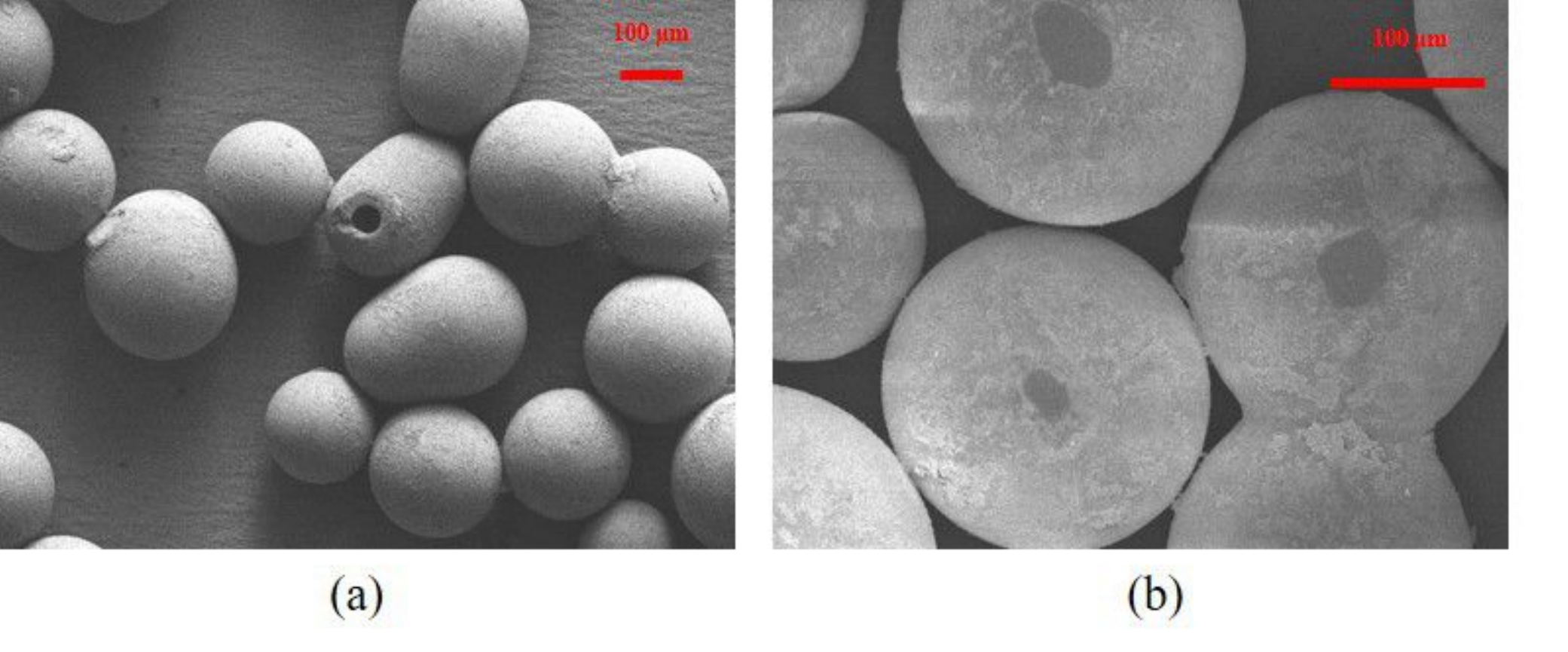}
	\centering
	\caption{Electron micrographs of acrylic copolymer (ACP) at (a) 100X (low magnification mode) and (b) 250X (secondary electron imaging).}
	\label{fig:SEMcopolymercoatedsurfaces-2}
\end{figure}

\begin{figure}[tbp]
	\includegraphics[width=0.7\textwidth]{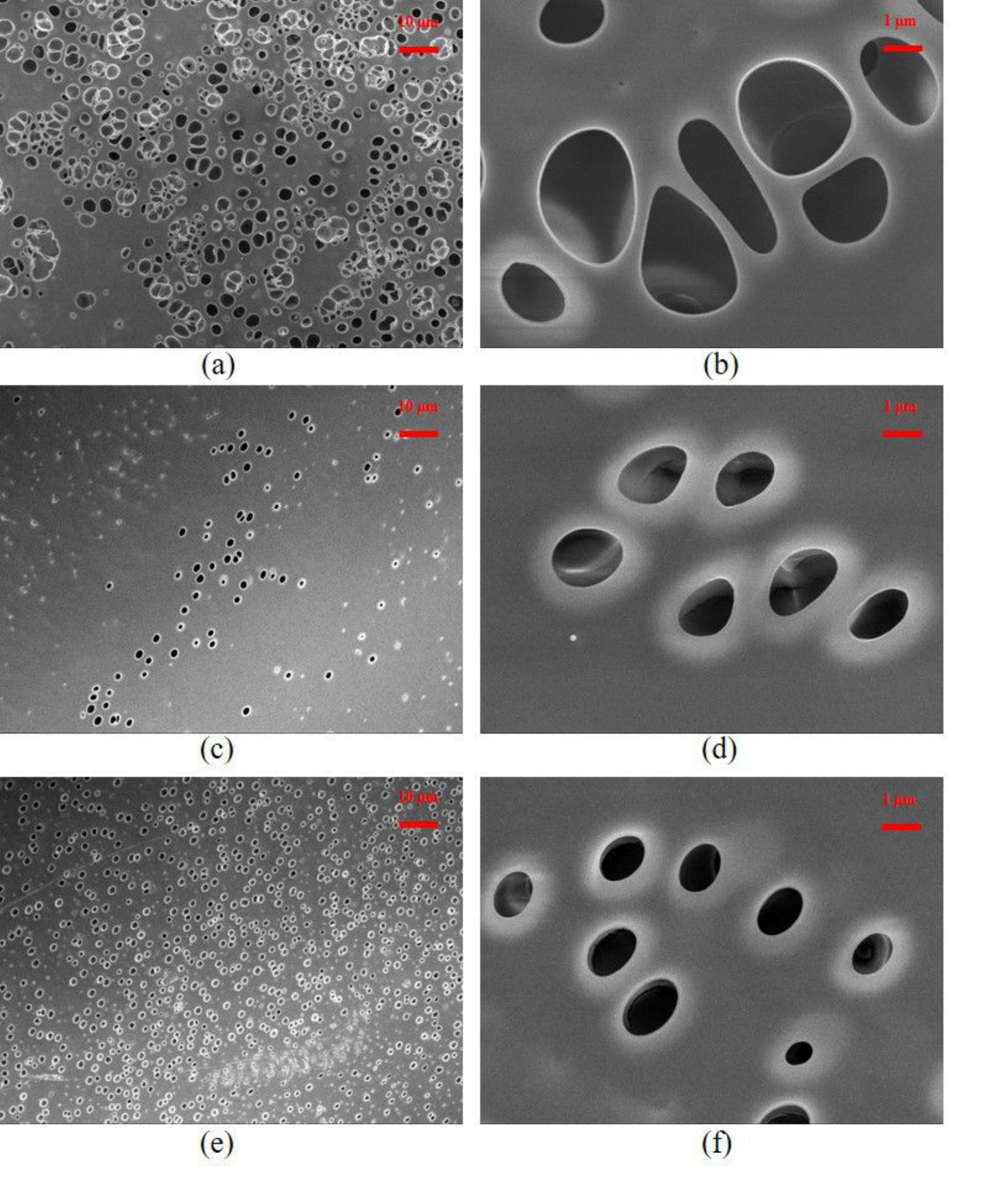}
	\centering
	\caption{Electron micrographs of acrylic copolymer (ACP) at different concentrations (in w/v): (a), (b) for 5\%, (c), (d) for 10\% and (e), (f) for 15\% at magnification of 1000X and 10000X respectively.}
	\label{fig:SEMcopolymercoatedsurfaces-1}
\end{figure}

For a cation exchange process, adsorbent (or stationary phase) consisting of large number of negative charges was required. Highly hydrophilic surface negates the possibility of protein adsorption and any weak cation exchange between the surface and proteins (\cite{Hasan2018}, \cite{Corsaro2021}, \cite{Aasim2022}). Hence, a suitable copolymer that could adhere to the chromatography column walls was to be chosen without being highly hydrophilic (ca. moderately high wettability). The hydrophilic character of acrylic copolymer (ACP) increases with higher copolymer concentration. It was found that 10\% copolymer concentration is sufficient for its usage in protein adsorption experiments further (Figures \ref{fig:copolymercontactangle1} and \ref{fig:copolymercontactangle2}). 
Electron microscopy images of these functional surfaces show microporous region at all the three acrylic copolymer concentrations (5, 10 and 15\%(w/v)) (Figure \ref{fig:SEMcopolymercoatedsurfaces-1}). These micropores were more likely to occur as induced by careful selection of both the solvent and surface functionalization scheme (Figure \ref{fig:copolymersurfacechemistryschematic}). Cracked surfaces could be obtained due to uncontrolled rate of solvent evaporation and similar observations have been reported in previous literature \cite{Hosseini2014}. The exception in that case was observed in in-house synthesis of acrylic copolymer as compared to the commercial product (supplied in the form of white beads at a different monomer ratio, see Figure \ref{fig:SEMcopolymercoatedsurfaces-2}) used in the present work. 

\subsection{Surface chemistry}\label{subsec10}
\begin{figure}[tbp]
	\includegraphics[width=0.7\textwidth]{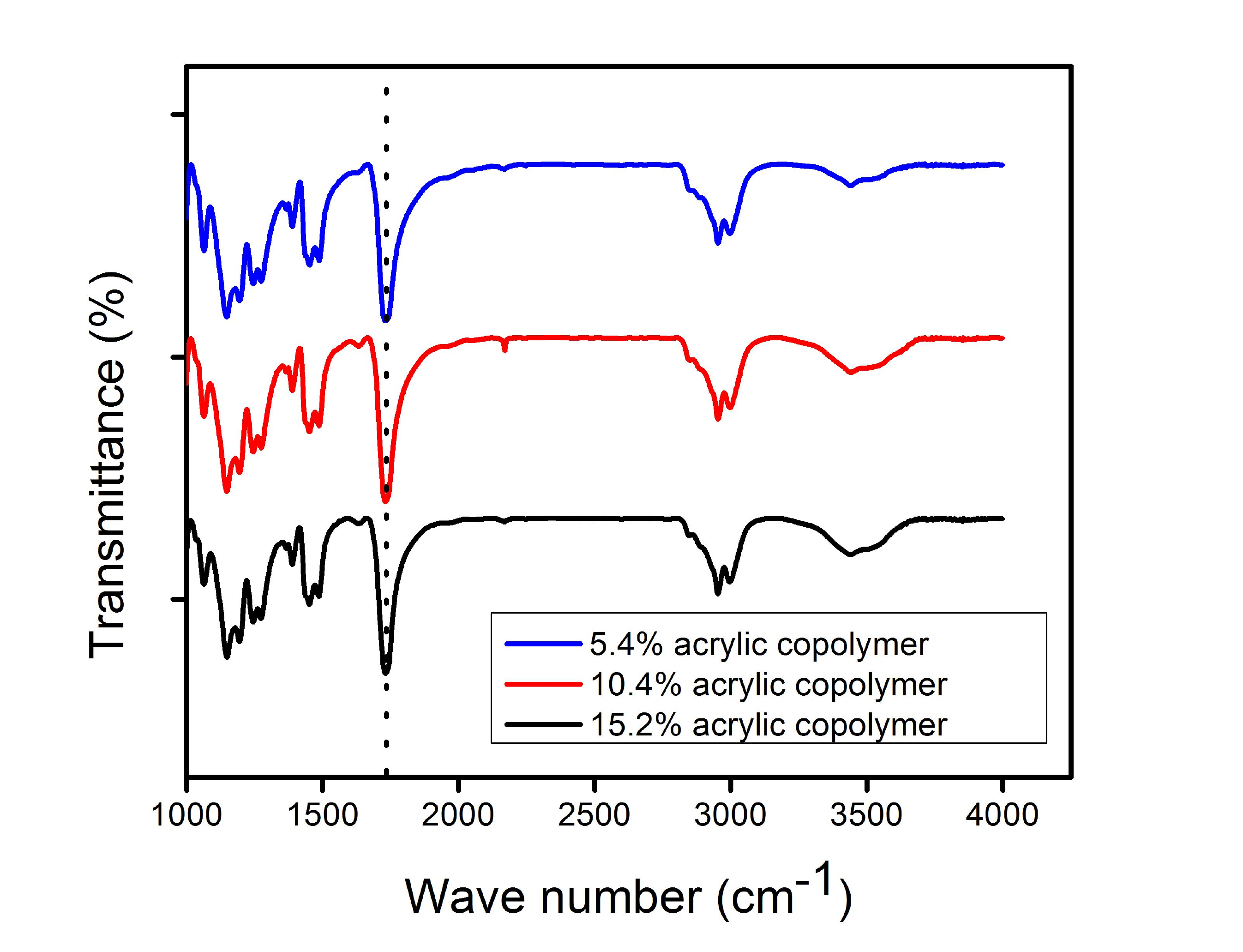}
	\centering
	\caption{FTIR spectrum of different concentrations of acrylic copolymer. Note that the transmittance with offset have been plotted on y-axis for display of stacked curves.}
	\label{fig:FTIRspectracopolymer}
\end{figure}

\begin{figure}[tbp]
	\includegraphics[width=0.7\textwidth]{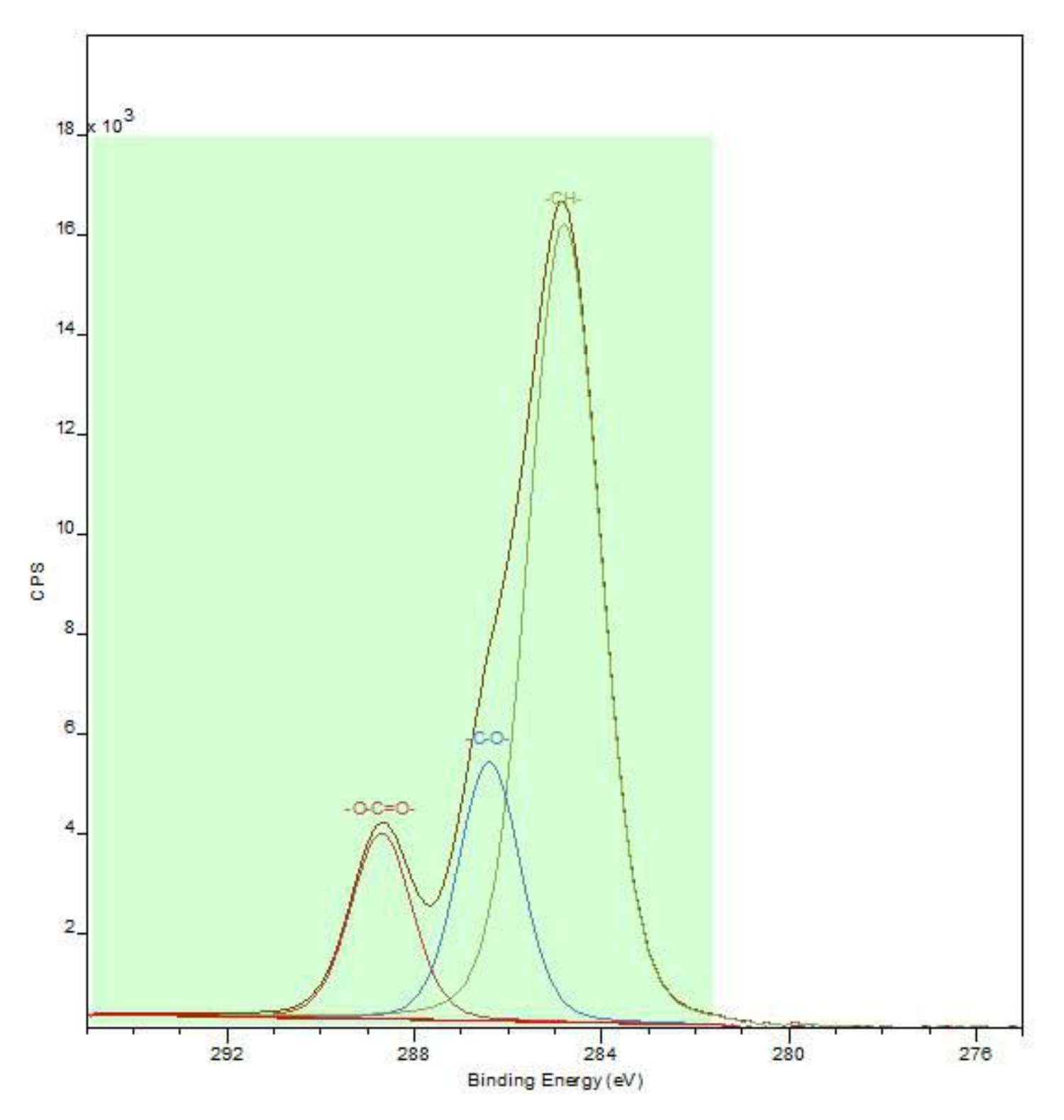}
	\centering
	\caption{X-ray photoelectron spectroscopy (XPS) of 10\% acrylic copolymer-modified PDMS surface.}
	\label{fig:XPSspectracopolymer}
\end{figure}

\begin{figure}[tbp]
	\includegraphics[width=0.7\textwidth]{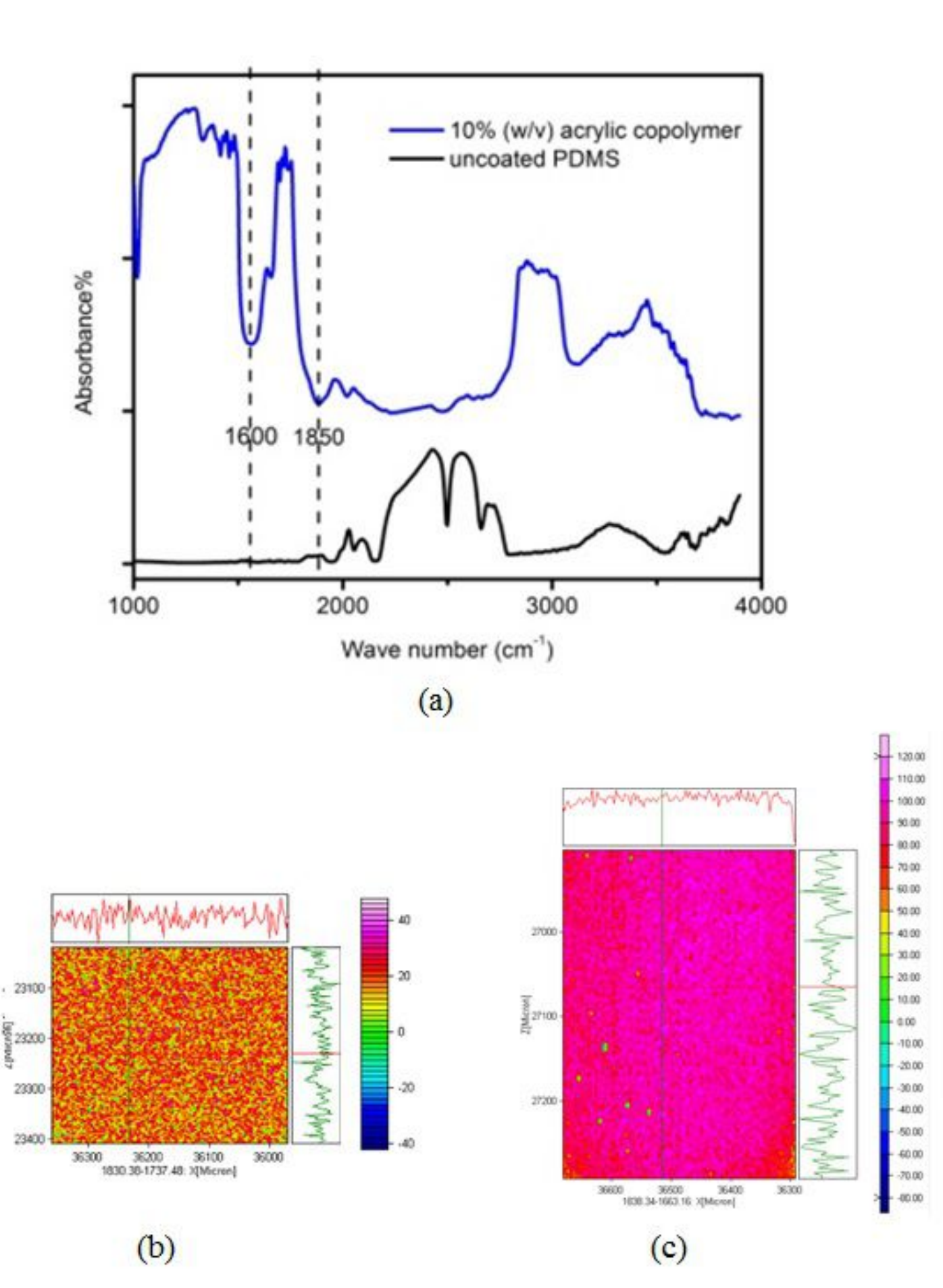}
	\centering
	\caption{FTIR analysis of different polymer surfaces: (a) FTIR spectral data for uncoated PDMS and 10\% (w/v) acrylic copolymer. Spectra were produced by grouping data into defined scan area of 128x128 pixels. Above FTIR spectra were an average of five different sampling points, and absorbance with offset have been plotted on y-axis for display of stacked curves. Chemical imaging  using FTIR imaging instrument with focal plane array (FPA) detector for (b) uncoated PDMS (b) 10\% (w/v) acrylic copolymer. Images (or chemical maps) were produced by grouping data into defined scan area of 128x128 pixels. Map tones indicate the integration intensity of carboxyl group (between 1600 to 1850 cm$^{-1}$). Pink tones correspond to high integration intensity confirming widespread presence of carboxyl groups.}
	\label{fig:FPAspectracopolymer}
\end{figure}

Two different blocks of acrylic copolymer (ACP), namely methyl methacrylate and methacrylic acid, facilitated as anchor for the functional surface on PDMS and covalently bonded carboxylate groups respectively. Analysis of these surfaces were carried out through contact angle, identification of functional groups of interest and corresponding chemical bonds. As part of functional material characterization, FTIR and XPS spectroscopy were utilized as analysis tools to confirm presence of negatively charged groups and carboxylate group (COO$^{-}$) in particular. Such chemical groups shall assist in improving the separation between various protein fractions during the ion exchange chromatography operation. The presence of carboxylic acid groups is confirmed by a sharp peak observed at 1731 cm$^{-1}$ for each acrylic copolymer concentration (see dotted line as shown in Figure \ref{fig:FTIRspectracopolymer}). The C1s-peak fitting analysis was carried out to confirm the binding energies of three different chemical bonds commonly identified with carboxyl groups (CASA XPS, USA). The reference peak of elemental carbon was considered at 284.6 eV (NIST, USA) for corrective purposes (Figure \ref{fig:XPSspectracopolymer}, where CPS - counts per second). The presence of chemical bonds with respective binding energies are as follows: O-C=O (288.7 eV), -C-O (286.4 eV), and -C-H- (284.8 eV); and that have been validated with previous literature \cite{Hosseini2014a}. The characteristic peak represents -C=O- bending vibration \cite{Duan2008} (Figure \ref{fig:FTIRspectracopolymer}) and an interval of wave numbers was chosen from the FTIR spectra collected over the functional surface (Figure \ref{fig:FPAspectracopolymer}). The copolymer surface inclusive of the porous region was found to have widespread presence of carboxylate groups. This confirmed the availability of specific binding sites with anions (in deprotonated state) suitable for weak cation exchange in protein chromatography.   

\begin{figure}[tbp]
	\includegraphics[width=0.9\textwidth]{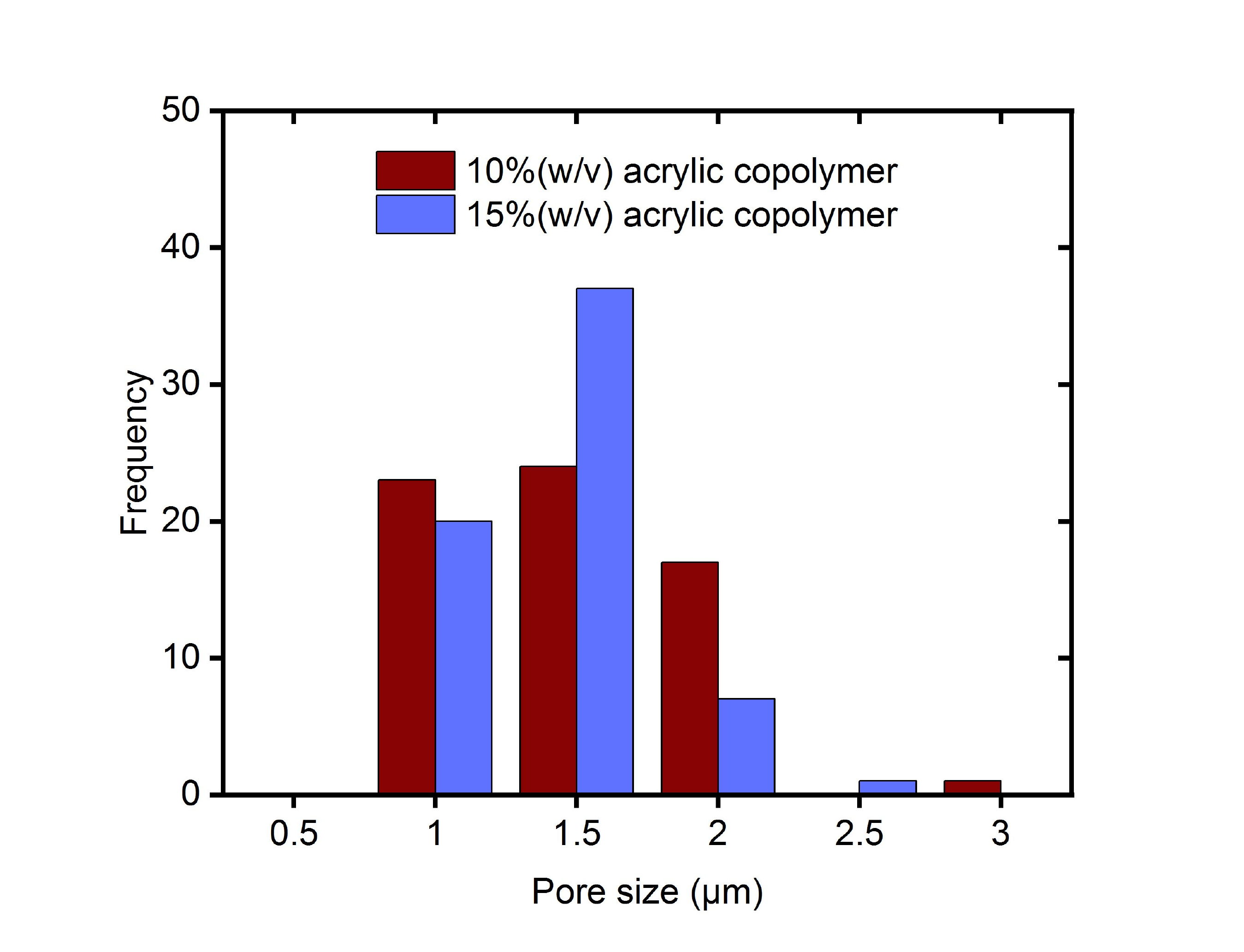}
	\centering
	\caption{Pore size distribution for micropores of 10\% (w/v) and 15\% (w/v) acrylic copolymer analyzed over duplicate set of scanning electron micrographs (a total of 66 pores analyzed in each image).}
	\label{fig:PSDcopolymer}
\end{figure}

Large pores relative to size of proteins indicate less hindrance to its diffusion likely towards the binding sites (\cite{Erickson2009}, \cite{Lebrun1994}). Trends in pore size distribution \cite{Abramoff2004} with varying acrylic copolymer concentration (5 to 15\% (w/v)) confirmed the presence of a microporous region (Figure \ref{fig:PSDcopolymer}). Size of micropores for 10\% ACP varied between 1 and 1.5 $\mu$m that were formed after near complete drying of solvent (THF). Effect of ACP concentration on the pore size is clearly evident from its size distribution and validates previous observations on formation of micropores (\cite{Kim2019}, \cite{Matsushita2004}) for a given combination of block copolymer and solvent (that consists of similar molecular weights).   

\subsection{Protein adsorption}\label{subsec11}
\begin{figure}[tbp]
	\includegraphics[width=0.9\textwidth]{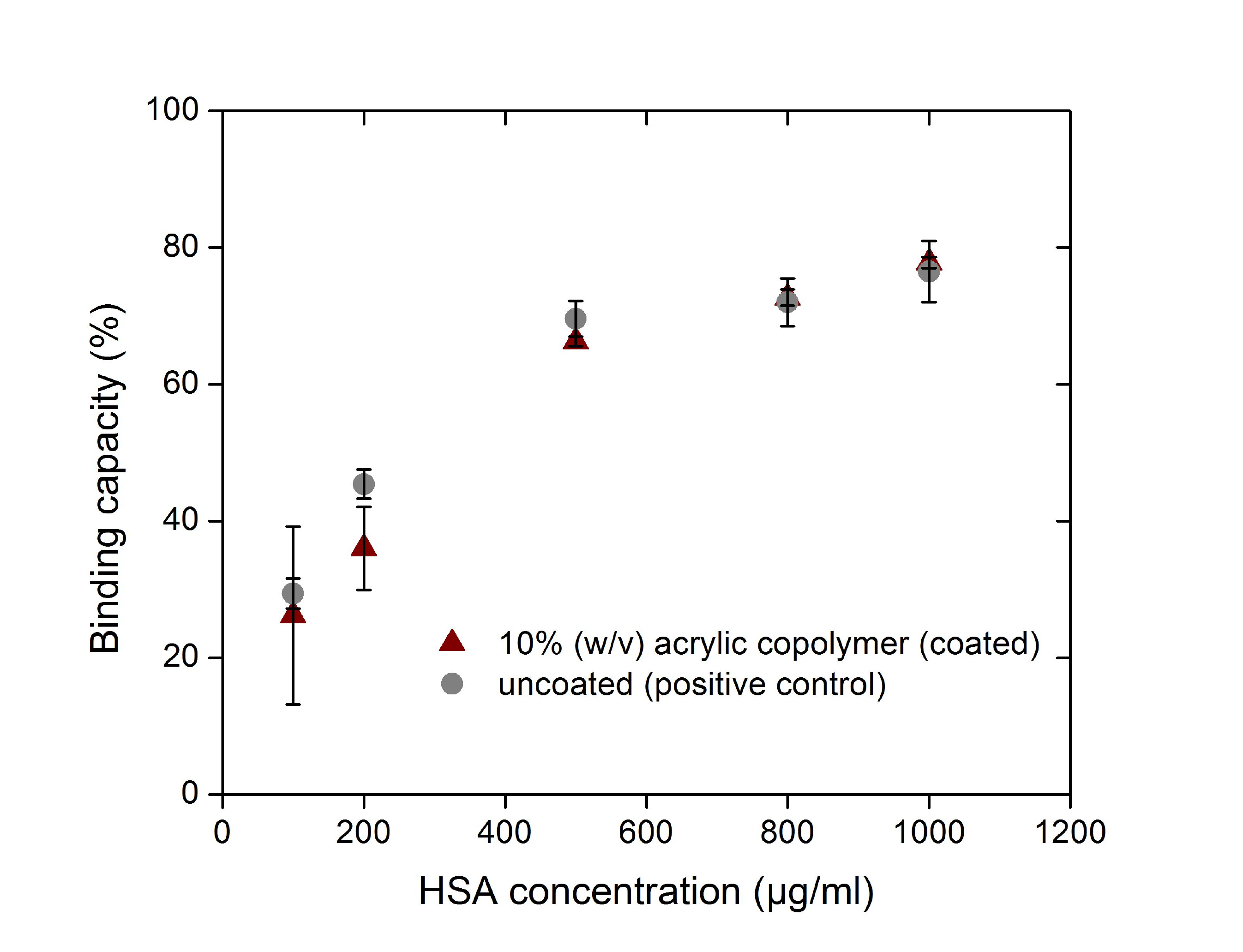}
	\centering
	\caption{Batch adsorption plot for uncoated and coated PDMS surface with different feed protein concentrations (100 to 1000 $\mu$g ml$^{-1}$). Feed protein: human serum albumin in phosphate buffer (pH 6.4), adsorbent: 10\% (w/v) acrylic copolymer (n=3)}
	\label{fig:batchbindingcapacity}
\end{figure}

The large functional pores in microporous region could allow proteins to diffuse into ACP (adsorbent).   
%Protein binding studies in replicates were performed and weight of protein analyzed using Bradford assay at microliter capacity. 
In batch mode, protein adsorption was carried out in comparison to uncoated PDMS surface in terms of binding capacity (Equation \ref{eqn:bindingcapacity}). For known feed protein concentration, similar binding capacities were obtained that indicate importance of ionic interactions between adsorbent and protein (Figure \ref{fig:batchbindingcapacity}). Porous ACP surface (or coated PDMS surface) exhibits similar binding capacity as uncoated PDMS surface when treated with a feed protein concentration of 500 $\mu$g ml$^{-1}$ or above. As peak resolution and separation is critical in protein chromatography \cite{Carta2020}, cation exchange media require these ionic interactions to be mediated by necessary functional groups such as carboxylic acid (carboxylate), sulfopropyl acid or sulfonic acid groups (\cite{Carta2020}, \cite{Hasan2018}, \cite{Phan2015}). Binding of proteins to the carboxylate groups provide control over bind-and-elute methods commonly employed in weak cation exchange chromatography. Uniform presence of these functional groups irrespective of how micropores are spread across the functional surface is an added advantage in case of ACP (Figures \ref{fig:SEMcopolymercoatedsurfaces-2} and \ref{fig:FPAspectracopolymer}). This has been validated with previously similar observations on adsorption of proteins from human plasma and serum samples reported by Sefton and co-workers \cite{Wells2017} except that porous ACP beads were in use. For packing protein chromatography columns, high pressure requirements required in the case of beads could be avoided through a shift to above coating technique along with the single-step scheme using functional copolymer surfaces. Hence, the emerging use case of open tubular LC columns \cite{Grinias2016} is likely to adopt wall-coated ACP as stationary phase instead of packed beads.

\section{Conclusion}\label{sec4}

Poly(methyl methacrylate-co-methacrylic acid) as microporous acrylic copolymer (ACP) is suitable as stationary phase for protein chromatography. Single-step scheme is implemented using ACP as functional copolymer surfaces along with uniform presence of functional groups. Without any additional functionalization steps, necessary protein adsorption characteristics such as hydrophilicity, cation exchange groups and corresponding chemical bonds are demonstrated in the present work. These characteristics are independent of ACP concentration and large pores that potentially facilitate access to protein binding sites. Thereby, these functional surfaces are likely to be of interest in bringing up innovative form factors of polymeric stationary phases such as packed monoliths and in other types of protein biosensors developed as part of bioanalytical techniques in the near future.

\backmatter

%\bmhead{Supplementary information}

%If your article has accompanying supplementary file/s please state so here. 

%Supplementary data on pore size distribution of different acrylic copolymer (or ACP) samples and data sheet on binding capacity of 10\% (w/v) ACP are included in this manuscript.

\bmhead{Acknowledgments}

We are thankful to the instrument facilities provided by Central Surface Analytical Facility (Department of Physics), Material Characterization and Testing Lab (Department of Chemical Engineering) and Sophisticated Analytical Instrumentation Facility (SAIF), at IIT Bombay.

\section*{Declarations}

%Some journals require declarations to be submitted in a standardised format. Please check the Instructions for Authors of the journal to which you are submitting to see if you need to complete this section. If yes, your manuscript must contain the following sections under the heading `Declarations':

\begin{itemize}
\item Funding: \\
One of the co-authors (Raghu K. Moorthy) duly acknowledges financial support received from Department of Science and Technology (DST), Government of India. 

\item Conflict of interest:\\
On behalf of all authors, the corresponding author states that there is no conflict of interest.

\item Ethics approval: \\
Not applicable

\item Consent to participate:\\
Not applicable

\item Consent for publication: \\
Not applicable

\item Availability of data and materials: \\
The datasets generated and analyzed during the current study are available from the corresponding author on reasonable request.

\item Code availability: 
Not applicable

\item Authors' contributions: 
Raghu K. Moorthy: Conceptualization, methodology, validation, formal analysis, investigation, data curation, writing - original draft, visualization. Serena D'Souza: Conceptualization, methodology, validation, formal analysis, data curation, writing - review \& editing, visualization, supervision. P. Sunthar: Conceptualization, methodology, formal analysis, resources, data curation, writing - review \& editing, visualization, supervision, project administration. Santosh B. Noronha:  Conceptualization, methodology, formal analysis, resources, writing - review \& editing, visualization, supervision, project administration, funding acquisition.

\end{itemize}

\noindent

\bibliography{sn-moorthy-et-al-22}% common bib file

\begin{thebibliography}{10}

\bibitem{Barrett2005}
D.~A. Barrett, G.~M. Power, M.~A. Hussain, I.~D. Pitfield, P.~N. Shaw, and
  M.~C. Davies.
\newblock {Protein interactions with model chromatographic stationary phases
  constructed using self-assembled monolayers}.
\newblock {\em J. Sep. Sci.}, 28:483--491, 2005.

\bibitem{Bratek-Skicki2019}
A.~Bratek-Skicki, V.~Cristaudo, J.~Savocco, S.~Nootens, P.~Morsomme,
  A.~Delcorte, and C.~Dupont-Gillain.
\newblock {Mixed polymer brushes for the selective capture and release of
  proteins}.
\newblock {\em Biomacromolecules}, 20(2):778--789, 2019.

\bibitem{Kang1993}
I.~Kang, B.~K. Kwon, J.~H. Lee, and H.~B. Lee.
\newblock {Immobilization of proteins on poly(methyl methacrylate) films}.
\newblock {\em Biomaterials}, 14(10):787--792, 1993.

\bibitem{Liu2000}
Y.~Liu, J.~C. Fanguy, J.~M. Bledsoe, and C.~S. Henry.
\newblock {Dynamic coating using polyelectrolyte multilayers for chemical
  control of electroosmotic flow in capillary electrophoresis microchips}.
\newblock {\em Anal. Chem.}, 72(24):5939--5944, 2000.

\bibitem{Becker2002}
H.~Becker and L.~E. Locascio.
\newblock {Polymer microfluidic devices}.
\newblock {\em Talanta}, 56(2):267--287, 2002.

\bibitem{Lee2010}
S.~H. Lee, D.~H. Kang, H.~N. Kim, and K.~Y. Suh.
\newblock {Use of directly molded poly(methyl methacrylate) channels for
  microfluidic applications}.
\newblock {\em Lab Chip}, 10(23):3300, 2010.

\bibitem{Ouimet2022}
J.~A. Ouimet, J.~Xu, C.~Flores‐Hansen, W.~A. Phillip, and B.~W. Boudouris.
\newblock {Design considerations for next‐generation polymer sorbents: From
  polymer chemistry to device configurations}.
\newblock {\em Macromol. Chem. Phys.}, page 2200032, 2022.

\bibitem{Kenney1968}
J.~F. Kenney.
\newblock {Properties of block versus random copolymers}.
\newblock {\em Polym. Eng. Sci.}, 8(3):216--226, 1968.

\bibitem{Hogt1985}
A.~H. Hogt, D.~E. Gregonis, I.~J.~D. Andrade, S.~W. Kim, J.~Dankert, and
  J.~Feijen.
\newblock {Wettability and $\zeta$ potentials of a series of methacrylate
  polymers and copolymers}.
\newblock {\em J. Colloid Interface Sci.}, 106(2):289--298, 1985.

\bibitem{Saunders1997}
B.~R. Saunders, H.~M. Crowther, and B.~Vincent.
\newblock {Poly[(methyl methacrylate)- co -(methacrylic acid)] microgel
  particles: Swelling control using pH, cononsolvency, and osmotic deswelling}.
\newblock {\em Macromolecules}, 30:482--487, 1997.

\bibitem{Pulker1999}
H.~K. Pulker.
\newblock {\em {Coatings on Glass}}.
\newblock Elsevier, second edition, 1999.

\bibitem{Sigma2018}
{Merck Inc. (formerly Sigma-Aldrich), USA, Technical notes (Glass transition
  temperature data sheet for polymers, Cat. no. 182230 and 376914}, 2018.
\newblock a US-based multi-national company in the field of chemicals and
  related products. Web link : http://sigmaaldrich.com/ Accessed online : 8 May
  2018.

\bibitem{Carta2020}
G.~Carta and A.~Jungbauer.
\newblock {\em {Protein Chromatography: Process Development and Scale-Up}}.
\newblock Wiley-VCH, 2 edition, 2020.

\bibitem{Hasan2018}
A.~Hasan, G.~Waibhaw, and L.~M. Pandey.
\newblock {Conformational and organizational insights into serum proteins
  during competitive adsorption on self-assembled monolayers}.
\newblock {\em Langmuir}, 34(28):8178--8194, 2018.

\bibitem{Phan2015}
H.~T.~M. Phan, S.~Bartelt-Hunt, K.~B. Rodenhausen, M.~Schubert, and J.~C.
  Bartz.
\newblock {Investigation of bovine serum albumin (BSA) attachment onto
  self-assembled monolayers (SAMs) using combinatorial quartz crystal
  microbalance with dissipation (QCM-D) and spectroscopic ellipsometry (SE)}.
\newblock {\em PLoS One}, 10:e0141282, 2015.

\bibitem{Hjerten1985}
S.~Hjert{\'{e}}n.
\newblock {High-performance electrophoresis. Elimination of electroendosmosis
  and solute adsorption}.
\newblock {\em J. Chromatogr. A}, 347:191--198, 1985.

\bibitem{Hoek2010}
I.~Hoek and W~M. Arnold.
\newblock {Sodium hydroxide treatment of PDMS based microfluidic devices}.
\newblock {\em Lab Chip}, 10:2283--2285, 2010.

\bibitem{Hosseini2014}
S.~Hosseini, F.~Ibrahim, I.~Djordjevic, and L.~H. Koole.
\newblock {Polymethyl methacrylate-co-methacrylic acid coatings with
  controllable concentration of surface carboxyl groups: A novel approach in
  fabrication of polymeric platforms for potential bio-diagnostic devices}.
\newblock {\em Appl. Surf. Sci}, 300:43--50, 2014.

\bibitem{Hosseini2014a}
S.~Hosseini, F.~Ibrahim, I.~Djordjevic, H.~A Rothan, R.~Yusof, C.~van~der
  Marel, and L.~H. Koole.
\newblock {Synthesis and processing of ELISA polymer substitute: The influence
  of surface chemistry and morphology on detection sensitivity}.
\newblock {\em Appl. Surf. Sci.}, 317:630--638, 2014.

\bibitem{Duan2008}
G.~Duan, C.~Zhang, A.~Li, X.~Yang, L.~Lu, and X.~Wang.
\newblock {Preparation and characterization of mesoporous zirconia made by
  using a poly (methyl methacrylate) template}.
\newblock {\em Nanoscale Res. Lett.}, 3(3):118--122, 2008.

\bibitem{Bradford1976}
M.~M. Bradford.
\newblock {A rapid and sensitive method for the quantitation of microgram
  quantities of protein utilizing the principle of protein-dye binding}.
\newblock {\em Anal. Biochem.}, 72(1-2):248--254, 1976.

\bibitem{Grintzalis2015}
K.~Grintzalis, C.~D. Georgiou, and Y.~J. Schneider.
\newblock {An accurate and sensitive Coomassie Brilliant Blue G-250-based assay
  for protein determination}.
\newblock {\em Anal. Biochem.}, 480:28--30, 2015.

\bibitem{Corsaro2021}
C.~Corsaro, D.~Mallamace, G.~Neri, and E.~Fazio.
\newblock {Hydrophilicity and hydrophobicity: Key aspects for biomedical and
  technological purposes}.
\newblock {\em Phys. A: Stat. Mech. Appl.}, 580:126189, 2021.

\bibitem{Aasim2022}
M.~Aasim, M.~H. Khan, N.~S. Bibi, and M.~Fernandez-Lahore.
\newblock {Understanding the interaction of proteins to ion exchange
  chromatographic supports: A surface energetics approach}.
\newblock {\em Biotechnology Progress}, 38:1--11, 2022.

\bibitem{Erickson2009}
H.~P. Erickson.
\newblock {Size and shape of protein molecules at the nanometer level
  determined by sedimentation, gel filtration, and electron microscopy}.
\newblock {\em Biol. Proced. Online}, 11:32--51, 2009.

\bibitem{Lebrun1994}
L.~Lebrun and G.~A. Junter.
\newblock {Diffusion of dextran through microporous membrane filters}.
\newblock {\em J. Memb. Sci.}, 88:253--261, 1994.

\bibitem{Abramoff2004}
M.~D. Abr{\`a}moff, P.~J. Magalh{\~a}es, and S.~J. Ram.
\newblock Image processing with imagej.
\newblock {\em Biophotonics Int.}, 11(7):36--42, 2004.

\bibitem{Kim2019}
I.~Kim and S.~Li.
\newblock {Recent progress on polydispersity effects on block copolymer phase
  behavior}.
\newblock {\em Polym. Rev.}, 59(3):561--587, 2019.

\bibitem{Matsushita2004}
A.~Noro, M.~Iinuma, J.~Suzuki, A.~Takano, and Y.~Matsushita.
\newblock {Effect of composition distribution on microphase-separated structure
  from diblock copolymers}.
\newblock {\em Macromolecules}, 37(10):3804--3808, 2004.

\bibitem{Wells2017}
L.~A. Wells, H.~Guo, A.~Emili, and M.~V. Sefton.
\newblock {The profile of adsorbed plasma and serum proteins on methacrylic
  acid copolymer beads: Effect on complement activation}.
\newblock {\em Biomaterials}, 118:74--83, 2017.

\bibitem{Grinias2016}
J.~P. Grinias and R.~T. Kennedy.
\newblock {Advances in and prospects of microchip liquid chromatography}.
\newblock {\em TrAC - Trends Anal. Chem.}, 81:110--117, 2016.

\bibitem{Aasim2012}
M.~Aasim.
\newblock {\em Surface Energetics of Protein Adsorption onto Chromatographic
  Supports}.
\newblock PhD thesis, Jacobs University, Bremen, 2012.

\end{thebibliography}
\bibliographystyle{unsrt}

\begin{table}[b]
	\setlength\belowcaptionskip{5pt}
	\centering
	\caption{Comparison of relevant physical properties}
	\label{tab:substratecomparison}
	%	\resizebox{\columnwidth}{!}{%
	\begin{tabular}{p{1cm}lllll}
		\hline
		{Material} & {Specific gravity*} &  \multicolumn{3}{c}{Solubility at 25$^{o}$C*}\\ 
		\cline{3-5}
		&  & Water & Toluene & Tetrahydrofuran  \\ \hline
		PDMS & 1.03  & Insoluble & High swelling & High swelling  \\
		& (on curing) & & & \\
		MAA & 1.015 & Soluble (HT) & Insoluble & Poor swelling \\
		PMMA & 1.188 & Insoluble & Soluble (HT) & Soluble\\
		ACP & 1.06 & Insoluble & NA & Soluble\\
		\hline
	\end{tabular}
	%	HT - at higher temperatures
	\begin{flushleft}
		*\cite{Pulker1999}, \cite{Sigma2018} \\ 
		PDMS - poly(dimethylsiloxane); MAA - methacrylic acid; PMMA - poly(methyl methacrylate); ACP - poly(methyl methacrylate-co-methacrylic acid); HT - at higher temperatures; NA - not available \end{flushleft}
\end{table}

\begin{table}[b]
	\setlength\belowcaptionskip{5pt}
	\centering
	\caption{Comparison of functionalized surfaces and functional surfaces}
	\label{tab:functionalizationliterature}
	%	\resizebox{\columnwidth}{!}{%
	\begin{tabular}{p{1cm}lllll}
		\hline
		{Material} & {Additional step(s)} & {Water contact angle} & {Reference} \\
		{} & {for surface functionalization} & {in Degrees} & \\ \hline
		PMMA & No & 90 & \cite{Hogt1985} \\
		PMMA$^{\mathrm{a}}$ & No & 90 or above & \cite{Aasim2022},\cite{Aasim2012} \\
		PMMA$^{\mathrm{a}}$ & Yes$^{\mathrm{b}}$ & 0 to 10 & \cite{Aasim2022},\cite{Aasim2012} \\
		ACP & No & 72 & \cite{Hosseini2014} \\
		ACP & No & 61 & Present work \\
		\hline
	\end{tabular}
	%	HT - at higher temperatures
	\begin{flushleft}
		$^{\mathrm{a}}$Polymethacrylates are the common base matrices for several commercial stationary phases (\cite{Aasim2022},\cite{Aasim2012}) \\
		$^{\mathrm{b}}$Surface functionalized with ether and(or) hydroxyl groups (\cite{Aasim2022},\cite{Aasim2012}) \\
		PMMA - poly(methyl methacrylate); ACP - poly(methyl methacrylate-co-methacrylic acid) \end{flushleft}
\end{table}

\end{document}